\UseRawInputEncoding
\documentclass[aps,twocolumn,prd,showpacs,showkeys,preprintnumbers,superscriptaddress,nobibnotes,floatfix,longbibliography,nofootinbib]{revtex4-1}

\pdfoutput=1
\usepackage{amsmath}
\usepackage{amsfonts}
\usepackage{amssymb}
\usepackage{mathrsfs}
\usepackage{graphicx}
\usepackage{color}
\usepackage{hyperref}
\usepackage{multirow}
\usepackage{slashed}
\usepackage{epsf}
\usepackage{amsmath,amssymb}
\usepackage{subcaption}
\usepackage{color}

\def\to{\rightarrow}

\begin{document}

\title{
The Natural Explanation of the Muon Anomalous Magnetic Moment via the Electroweak Supersymmetry
from the GmSUGRA in the MSSM
}

\author{Waqas Ahmed}
\affiliation{School of Physics, Nankai University, No.94 Weijin Road, Nankai District, Tianjin, China}

\author{Imtiaz Khan}
\affiliation{CAS Key Laboratory of Theoretical Physics, Institute of Theoretical Physics, Chinese Academy of Sciences, Beijing 100190, China}
\affiliation{School of Physical Sciences, University of Chinese Academy of Sciences, No. 19A Yuquan Road, Beijing 100049, China}

\author{Jinmian Li}
\affiliation{College of Physics, 
	Sichuan University, 
	Chengdu 610065, China}

\author{Tianjun Li}
\affiliation{CAS Key Laboratory of Theoretical Physics, Institute of Theoretical Physics, Chinese Academy of Sciences, Beijing 100190, China}
\affiliation{School of Physical Sciences, University of Chinese Academy of Sciences, No. 19A Yuquan Road, Beijing 100049, China}

\author{Shabbar Raza}
\affiliation{Department of Physics, Federal Urdu University of Arts, Science and Technology, Karachi 75300, Pakistan}
\author{Wenxing Zhang}
\affiliation{Tsung-Dao Lee Institute and School of Physics and Astronomy, Shanghai Jiao Tong University, 800 Dongchuan Rd., 
Minhang, Shanghai 200240, China}



\begin{abstract}

The Fermi-Lab Collaboration has announced the results for the measurement of
 muon anomalous magnetic moment. Combining with the previous results
by the BNL experiment, we have $4.2 \sigma$ deviation
from the Standard Model (SM), which strongly implies the new physics around 1 TeV. To explain the 
muon anomalous magnetic moment naturally, we analyze the corresponding five Feynman diagrams
in the Supersymetric SMs (SSMs), and show that the Electroweak Supersymmetry (EWSUSY) is definitely 
needed. We realize the EWSUSY in the Minimal SSM (MSSM) with Genernalized Mininal 
Supergravity (GmSUGRA). We find large viable parameter space, which is consistent
with all the current experimental constraints. In particular, the Lightest 
Supersymmetric Particle (LSP) neutralino can be at least as heavy as 550 GeV.
Most of the viable parameter space can be probed at the future HL-LHC,
while we do need the future HE-LHC to probe some viable parameter space.  
However, it might still be challenge if R-parity is violated.

\end{abstract}
\maketitle

\textbf{Introduction.--} After Higgs particle was discovered at the LHC, the Standard Model (SM)
has been confirmed. However, there exists some problems, and thus we need to 
study the new physics beyond the SM. It is well-known that Supersymmetry (SUSY) is
the most natural solution to the gauge hierarchy problem.
In the supersymmetric SMs (SSMs) with R-parity,
 we can achieve gauge coupling unification~\cite{gaugeunification}, have 
the Lightest Supersymmetric Particle (LSP) like the lightest neutralino 
as dark matter (DM) candidate~\cite{Jungman:1995df},
and break the  electroweak  (EW)  gauge symmetry radiatively because of 
the large top quark Yukawa coupling, etc.
Moreover, gauge coupling unification strongly implies the
Grand Unified Theories (GUTs)~\cite{Georgi:1974sy,Pati:1974yy,Mohapatra:1974hk,Fritzsch:1974nn,Georgi:1974my}, 
and the SSMs and SUSY GUTs can be constructed from superstring theory.  
Therefore, supersymmetry provides a bridge between 
the low energy phenomenology and high-energy fundamental physics, and thus is 
the most promising new physics beyond the SM.

However, after the second run at the Large Hadron Collider (LHC), 
we still did not have any SUSY signals, and then
the LHC SUSY searches have already given strong constraints on the SSMs. For instance, 
 the masses of the gluino, first-two generation squarks, stop, and sbottom must be larger
than about 2.3~TeV, 1.9~TeV, 1.25~TeV, and 1.5~TeV, 
respectively~\cite{ATLAS-SUSY-Search, Aad:2020sgw, Aad:2019pfy, CMS-SUSY-Search-I, CMS-SUSY-Search-II}.
Thus, at least the colored supersymmetric particles (sparticles) must be heavy around a few TeV.

Interestingly,  a well-known long-standing deviation is
 a 3.7 $\sigma$  discrepancy for the muon anomalous magnetic moment $a_\mu = (g_\mu-2)/2$
between the experimental results~\cite{Bennett:2006fi, Tanabashi:2018oca} and 
theoretical predictions~\cite{Davier:2017zfy, Blum:2018mom, Keshavarzi:2018mgv, Davier:2019can}
\begin{equation}
\label{muon} \Delta a_\mu = a_\mu^{exp}-a_\mu^{th} = (27.4\pm7.3)\times 10^{-10}~.~\,
\end{equation} 
Computing the hadronic light-by-light contribution with all errors under control 
by using lattice QCD, several groups have tried to improve the precision of 
the SM predictions~\cite{Aubin:2019usy, Blum:2015you, Lehner:2019wvv, Davies:2019efs}.
And the $\Delta a_\mu$ discrepancy has been confirmed by the 
recent lattice calculation for  the hadronic light-by-light scattering contribution~\cite{Blum:2019ugy}, 
and then a new physics explanation is needed.
Also, the ongoing experiment at Fermilab~\cite{Grange:2015fou, Fienberg:2019ddu} 
and one planned at J-PARC~\cite{Saito:2012zz} will try to reduce the uncertainty. 

To escape the LHC SUSY search constraints, 
explain the muon anomalous magnetic moment, and be consistent with various experimental results,
some of us proposed 
the Electroweak Supersymmetry (EWSUSY)~\cite{Cheng:2012np, Cheng:2013hna, Li:2014dna}, 
where the squarks and/or gluinos are
around a few TeV while the sleptons, sneutrinos, Bino and Winos are within about 1 TeV. The
Higgsinos (or say the Higgs bilinear $\mu$ term) can be either heavy or light. Especially, 
the EWSUSY can be realized 
in the Generalized Minimal Supergravity (GmSUGRA)~\cite{Li:2010xr, Balazs:2010ha}.

Recently, the Fermi-Lab Collaboration has announced the results for the measurement of
 the anomalous magnetic moment of the muon. 
Combining with the previous results by the Brookhaven National Lab (BNL) experiment, 
we have $4.2 \sigma$ deviation from the SM~\cite{Fermi-Lab}
\begin{equation}
\label{muon} \Delta a_\mu = a_\mu^{exp}-a_\mu^{th} = (25.1\pm5.9)\times 10^{-10}~.~\,
\end{equation} 
Because it strongly suggests the new physics around 1 TeV, we shall explain the muon 
anomalous magnetic moment in the SSMs in this paper. Here, we will not consider the SSMs where
the supersymmetry breaking soft terms are introduced at the low scale since such kind of
scenarios has more freedoms and might not be consistent with the GUTs and string models.
To explain the muon anomalous magnetic moment naturally, we analyze five corresponding Feynman diagrams,
and show that the EWSUSY is definitely needed. We realize
the EWSUSY in the Minimal SSM (MSSM)  with the GmSUGRA. 
We find large viable parameter space, which is consistent
with all the current experimental constraints. In particular, the LSP neutralino can 
be at least as heavy as 550 GeV. Most of the viable parameter
space can be probed by the future High Luminosity-LHC (HL-LHC),
while we do need the future High Enery-LHC (HE-LHC) with center-of-mass energy 27 TeV 
to probe some viable parameter space. 
 However, it may be a big challenge if R-parity is violated.
\begin{figure}[ht]
	\begin{subfigure}{.15\textwidth}
		\includegraphics[width=1.0\linewidth]{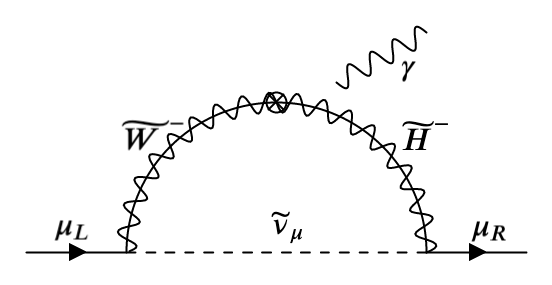}  
		\caption{\label{fig:fd-a}}
	\end{subfigure}
	\begin{subfigure}{.15\textwidth}
		\includegraphics[width=1.0\linewidth]{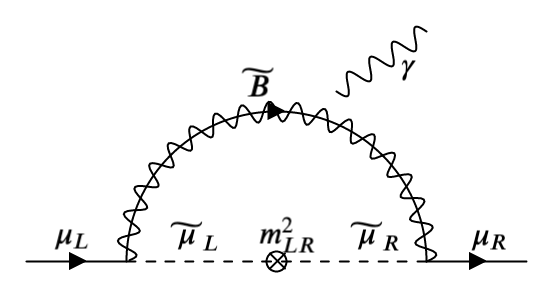}  
		\caption{\label{fig:fd-b}}
	\end{subfigure}
	\begin{subfigure}{.15\textwidth}
		\includegraphics[width=1.0\linewidth]{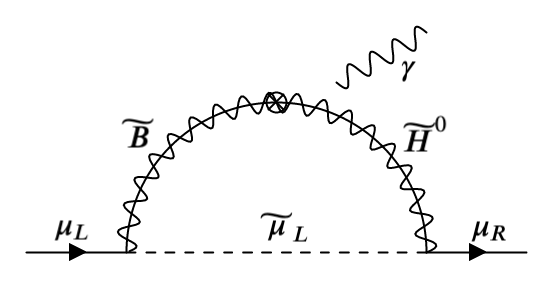}  
		\caption{\label{fig:fd-c}}
	\end{subfigure}
	\begin{subfigure}{.15\textwidth}
		\includegraphics[width=1.0\linewidth]{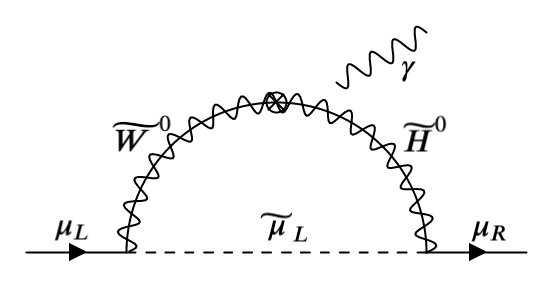}  
		\caption{\label{fig:fd-d}}
	\end{subfigure}
	\begin{subfigure}{.15\textwidth}
		\includegraphics[width=1.0\linewidth]{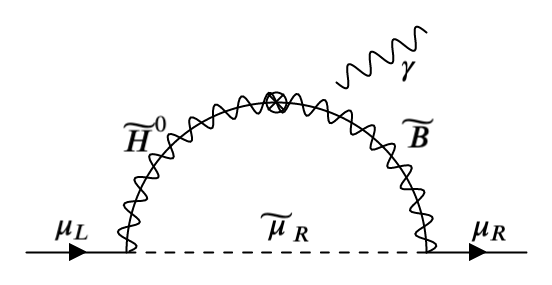}  
		\caption{\label{fig:fd-e}}
	\end{subfigure}
	\caption{Feynman diagrams that give the dominant SUSY contributions to the $a_\mu$.}
	\label{fig:feynman}
\end{figure}

\textbf{Muon Anomalous Magnetic Moment and EWSUSY.--} There are five Feynman diagrams 
in the SSMs which will contribute to  $\Delta a_\mu$~\cite{Cho:2011rk}, as given in
Fig. 1. First, for $M_2 \mu >0$ and $M_1 \mu >0$,
diagrams (a), (b), and (c) give positive contributions, while diagrams (d) and (e)
give negative contributions. Second, if Higgsino is very heavy,
only the diagram (b) will give dominant contribution. Third, if the mass splitting
between the muon sneutrino and left-handed smuon are small as in our study,
the sum of the diagrams (a) and (d) is positive, {\it i.e.}, the contributions from 
diagram (a) is generically larger. Fourth, the contribution from diagram (c) is always relatively 
smaller compared to these from diagrams (a) and (b) in our study. Of course, the contribution 
from diagram (c) can be dominant if we choose light Bino, Higgsino, and left-handed smuon, as well as
heavy $M_2$ and right-handed smuon by hand at low enery, and we have confirmed it numerically. 
However, this is not consistent with GUTs and string models since larger $M_2$ will increase 
the left-handed smuon mass due to the renormalization group equation (RGE) running. 
Fifth, we find that the contribution to $\Delta a_\mu$ from diagram (e) is smaller than
$6\times 10^{-10}$ in our study, and is generically smaller than $10\times 10^{-10}$, {\it i.e.}, out of $2\sigma$ region.
In short, within $2\sigma$ region, we can only explain the muon anomalous magnetic moment via 
diagrams (a) and (b). And then we obtain that
 the sleptons, sneutrinos, Bino and Winos must be light and cannot be much heavier
than 1 TeV. Also, if diagram (b) gives dominant contribution,
the Higgsinos can be very heavy, while Wino cannot be very heavy since it will contribute
to the left-handed smuon mass due to the RGE running. Therefore, we have shown that
EWSUSY is definitely needed to explain the muon anomalous magnetic moment.

\begin{figure}[t!]
\includegraphics[width=1.0\columnwidth]{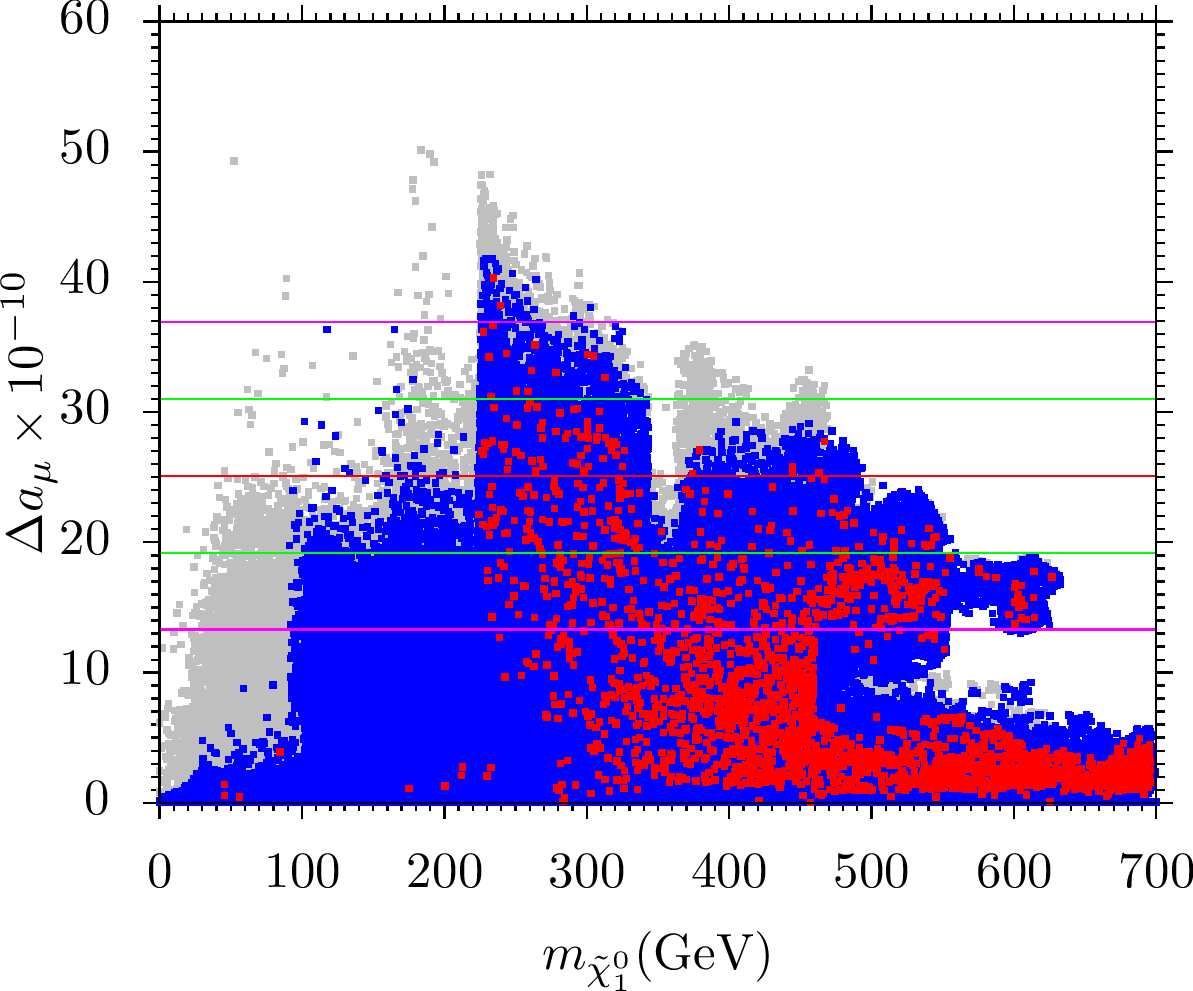}
\caption{
Gray points are consistent with the REWSB and LSP neutralino. Blue points satisfy the mass bounds including $m_{h}=125 \pm 2 \,{\rm GeV}$ and the constraints from rare $B-$ meson decays. Red points form a subset of blue points and satisfy the 5$\sigma$ Planck bounds on dark matter relic density. Red line shows the central value of $\Delta a_{\mu}$, and greens and purple lines represent 1$\sigma$ and 2$\sigma$ deviations from the central value.}
\label{fig:fig2}
\end{figure}


%
\begin{figure*}[ht]
    \centering
        \begin{tabular}{c c}
    \includegraphics[width = 0.5\textwidth]{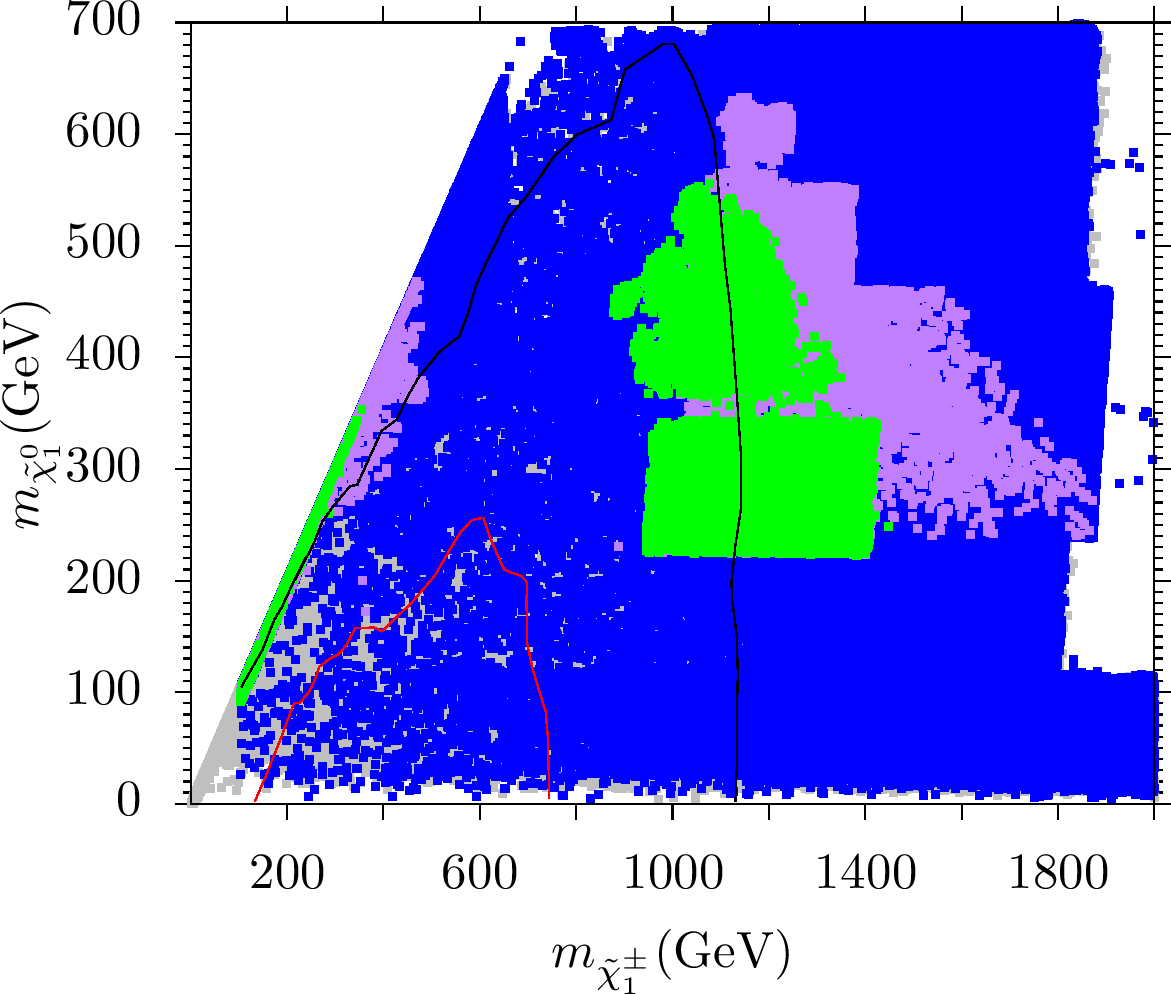} &
        \hspace{-.01cm}\includegraphics[width = 0.5\textwidth]{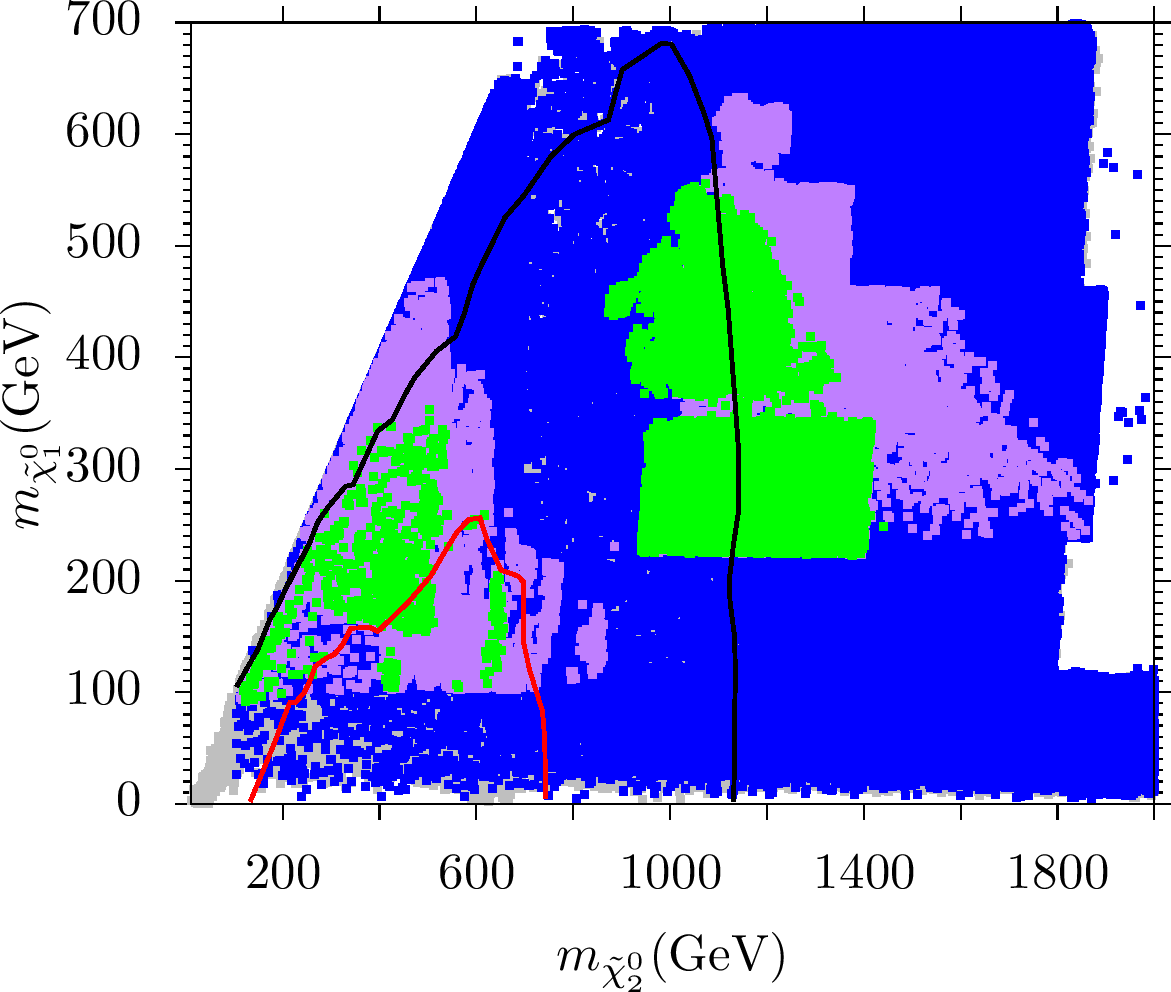}  \\
    \end{tabular}
    \caption{Gray points are consistent with the REWSB and LSP neutralino. Blue points satisfy the mass bounds including $m_{h}=125 \pm 2 \,{\rm GeV}$ and the constraints from rare $B-$ meson decays. 
The  green and purple points form the subsets of blue points and are consistent 
with 1$\sigma$ and 2 $\sigma$ deviations from the central value of $\Delta a_{\mu}$, respectively. 
Black and red curves represent the ATLAS SUSY search bounds \cite{ATLAS}}.
    \label{fig:fig3}
\end{figure*}




\textbf{The EWSUSY from the GmSUGRA in the MSSM.--} The EWSUSY can be realized in the GmSUGRA~\cite{Li:2010xr, Balazs:2010ha}, where the sleptons and electroweakinos (charginos, Bino, Wino, and/or Higgsinos) are within about 1 TeV while squarks and/or gluinos 
can be in several TeV mass ranges~\cite{Cheng:2012np, Cheng:2013hna, Li:2014dna}. 
The supersymmetry breaking soft (SBS) terms in the GmSUGRA~\cite{Li:2010xr, Balazs:2010ha} are
\begin{eqnarray}
M_3&=&\frac{5}{2}~M_1-\frac{3}{2}~M_2~, \\
m_{\tilde{Q}_i}^2 &=& \frac{5}{6} (m_0^{U})^2 +  \frac{1}{6} m_{\tilde {E}_i^c}^2~,\\
m_{\tilde {U}_i^c}^2 &=& \frac{5}{3}(m_0^{U})^2 -\frac{2}{3} m_{\tilde{E}_i^c}^2~,\\
m_{\tilde {D}_i^c}^2 &=& \frac{5}{3}(m_0^{U})^2 -\frac{2}{3} m_{\tilde {L}_i}^2~,
\label{squarks_masses}
\end{eqnarray}
where $M_1$, $M_2$, $M_3$ are the masses for Bino, Wino, and gluino at the GUT scale, respectively,
as well as $m_{\tilde {Q}}$, $m_{\tilde {U^c}}$, $m_{\tilde {D^c}}$, $m_{\tilde L}$, and  $m_{\tilde E^c}$ represent 
the scalar masses of the left-handed squark doublets, right-handed up-type squarks, right-handed down-type squarks,
left-handed sleptons, and right-handed sleptons, respectively, while $m_0^U$ is the universal  
scalar mass, as in the mSUGRA. In the EWSUSY, $m_{\tilde L}$ and $m_{\tilde E^c}$ are both within about 1 TeV, resulting in 
light sleptons. Especially, in the limit $m_0^U \gg m_{\tilde {L}/\tilde {E^c}}$, we have the approximated 
relations for squark masses: $2 m_{\tilde {Q}}^2 \sim m_{\tilde {U^c}}^2 \sim m_{\tilde {D^c}}^2$. In addition, 
the Higgs soft masses $m_{{H_u}}$ and $m_{{H_d}}$, and the  trilinear soft terms
 $A_U$, $A_D$ and $A_E$ can all be free parameters from the GmSUGRA~\cite{Cheng:2012np,Balazs:2010ha}.



\textbf{Scanning Process.--} We employ the ISAJET~7.85 package~\cite{ISAJET}
 to perform random scans over the parameter space
 given below.
In this package, the weak-scale values of the gauge and third
 generation Yukawa couplings are evolved to
 $M_{\rm GUT}$ via the MSSM RGEs
 in the $\overline{DR}$ regularization scheme.
We do not strictly enforce the unification condition
 $g_3=g_1=g_2$ at the GUT scale $M_{\rm GUT}$, since a few percent deviation
 from unification can be assigned to the unknown GUT-scale threshold
 corrections~\cite{Hisano:1992jj}.
With the boundary conditions given at $M_{\rm GUT}$,
 all the SBS parameters, along with the gauge and Yukawa couplings,
 are evolved back to the weak scale $M_{\rm Z}$ (for more detail see \cite{ISAJET}).
We have performed the random scans
 for the following parameter ranges
\begin{align}
100 \, \rm{GeV} \leq & m_0^{U}  \leq 10000 \, \rm{GeV}  ~,~\nonumber \\
100 \, \rm{GeV} \leq & |M_1|  \leq 1600 \, \rm{GeV} ~,~\nonumber \\
100\, \rm{GeV} \leq & |M_2|   \leq 10000 \, \rm{GeV} ~,~\nonumber \\
100 \, \rm{GeV} \leq & m_{\tilde L}  \leq 10000 \, \rm{GeV} ~,~\nonumber \\
100 \, \rm{GeV} \leq & m_{\tilde E^c}  \leq 1000 \, \rm{GeV} ~,~\nonumber \\
-10000 \, \rm{GeV} \leq & m_{\tilde H_{u,d}} \leq 10000 \, \rm{GeV} ~,~\nonumber \\
-10000 \, \rm{GeV} \leq & A_{U}=A_{D} \leq 10000 \, \rm{GeV} ~,~\nonumber \\
-10000 \, \rm{GeV} \leq & A_{E} \leq 10000 \, \rm{GeV} ~,~\nonumber \\
2\leq & \tan\beta  \leq 60~.~
 \label{input_param_range}
\end{align}
Also, we consider  $\mu > 0$, and  use $m_t = 173.3\, {\rm GeV}$ 
and $m_b^{\overline{DR}}(M_{\rm Z})=2.83$ GeV \cite{:2009ec}.
Note that our results are not too sensitive to one
 or two sigma variations in the value of $m_t$  \cite{bartol2}.
Also, we will use the notations $A_t,A_b,A_{\tau}$ for $A_{U},A_D$ and $A_E$, respectively.
 In scanning the parameter space, we employ the Metropolis-Hastings
 algorithm as described in \cite{Belanger:2009ti}.
{\it In particular, we also perform some focus scans inspired from the diagrams (a) and (b),
which do give us better viable parameter spaces to explain the $\Delta a_{\mu}$.}
The collected data points all satisfy the requirement of the 
 Radiative Electroweak Symmetry Breaking (REWSB),
 with the lightest neutralino being the LSP.

\textbf{Constraints.--} After collecting the data, we impose the bounds that the LEP2 experiments set on charged sparticle masses ($\gtrsim 100$ GeV) \cite{Patrignani:2016xqp}, for Higgs mass bounds \cite{Khachatryan:2016vau} 
 due to an estimated 2 GeV theoretical uncertainty in the calculation of $m_h$ in the MSSM -- see {\it e.g.}~\cite{Allanach:2004rh} --
we apply the constraint from the Higgs boson mass to our results as $m_{h}=[122,128] {\rm GeV}$. In addition, based on~\cite{ATLAS-SUSY-Search, Aad:2020sgw, Aad:2019pfy, CMS-SUSY-Search-I, CMS-SUSY-Search-II}, we consider the constraints on gluino $m_{\widetilde g} \gtrsim ~ 2.2 \,{\rm TeV}$. The constraints from rare decay processes $B_{s}\rightarrow \mu^{+}\mu^{-} $ \cite{Aaij:2012nna}, $b\rightarrow s \gamma$ \cite{Amhis:2012bh}, and $B_{u}\rightarrow \tau\nu_{\tau}$ \cite{Asner:2010qj}. We also require the relic abundance of the LSP neutralino to satisfy the Planck bound within $5\sigma$ \cite{Akrami:2018vks}. More explicitly, we set
\begin{eqnarray}
m_h  = 122-128~{\rm GeV}~~&
\\
m_{\tilde{g}}\geq 2.2~{\rm TeV} \\
0.8\times 10^{-9} \leq{\rm BR}(B_s \rightarrow \mu^+ \mu^-)
  \leq 6.2 \times10^{-9} \;(2\sigma)~~&&
\\
2.99 \times 10^{-4} \leq
  {\rm BR}(b \rightarrow s \gamma)
  \leq 3.87 \times 10^{-4} \; (2\sigma)~~&&
\\
0.15 \leq \frac{
 {\rm BR}(B_u\rightarrow\tau \nu_{\tau})_{\rm MSSM}}
 {{\rm BR}(B_u\rightarrow \tau \nu_{\tau})_{\rm SM}}
        \leq 2.41 \; (3\sigma)~~&&
\\
 0.114 \leq \Omega_{\rm CDM}h^2 (\rm Planck) \leq 0.126   \; (5\sigma)~~&&.
\end{eqnarray}

\textbf{Results.--} We shall discuss results of the scans in the following. 
In Fig.~\ref{fig:fig2}, we display plot in $m_{\tilde \chi_{1}^{0}}-\Delta a_{\mu}$ plane. 
So in our model, we can explain $\Delta a_{\mu}$ very easily. In particular,
some viable parameter spaces including the viable parameter spaces with
the correct dark matter relic density have $\Delta a_{\mu}$ around
the central value $25.1\times 10^{-10}$. Note that the peak at $m_{\tilde \chi_{1}^{0}}\sim 300 \,{\rm GeV}$ is not the unique feature but the artifact of focused scans. Another important point to be noted is that 
the solutions with relatively heavy LSP neutralino $\sim 600 \,{\rm GeV}$ are still consistent within 2$\sigma$ values of $\Delta a_{\mu}$. As far as we know, no one has found 
such kind of viable parameter space before. Thus, it is very interesting. However,
we want to make a comment here about the solutions with  $m_{\tilde \chi_{1}^{0}}> 600 \,{\rm GeV}$: 
 we have checked some points and found that some of them might not be numerically stable. 

In Fig.~\ref{fig:fig3}, we depict plots in $m_{\tilde \chi_{1}^{\pm}}-m_{\tilde \chi_{1}^{0}}$ and $m_{\tilde \chi_{2}^{0}}-m_{\tilde \chi_{1}^{0}}$. The solid black and red curves are the 95$\%$ CL exclusion limits on $\tilde \chi_{1}^{+}\tilde \chi_{1}^{-}$ and $\tilde \chi_{1}^{\pm}\tilde \chi_{2}^{0}$ from pair productions with $\tilde l$-mediated decays as a function of the $\tilde \chi_{1}^{\pm},\tilde \chi_{2}^{0}$ and $\tilde \chi_{1}^{0}$ masses \cite{ATLAS}. Purple and green patches as stated before are due to the focused scans. In the left panel, we see that we have two sets of solutions which satisfy $\Delta a_{\mu}$ within 1$\sigma$ and 2$\sigma$ that is the compressed region where $\tilde \chi_{1}^{0}$ and $\tilde \chi_{1}^{\pm}$ are almost degenerate in mass and region where $\tilde \chi_{1}^{0}$ and $\tilde \chi_{1}^{\pm}$ mass spliting is greater than $500\, {\rm GeV}$. Green solutions in the compressed region mostly belongs to diagram (a) of Fig.~\ref{fig:feynman} and the solutions with heavy chargino masses but light $m_{\tilde \chi_{1}^{0}}$ belongs to diagram (b) of Fig.~\ref{fig:feynman}. We show three benchmark points in Table.\ref{table1}. Point 1 represents a set of solutions with relatively heavy neutralino $m_{\tilde \chi_{1}^{0}}> 400 \,{\rm GeV}$. 
Because Wino and Higgsinos are heavy in Points 1 and 2, these points belong to the diagram (b) explanation.
 Point 3 is an example where Bino, Wino and sneutrino are light but Higgsinos are heavy, 
so diagrams (a) and (b) both contribute here.

\begin{table}[h!]
	\centering
	\scalebox{0.8}{
		\begin{tabular}{lccc}
			\hline
			\hline
			& Point 1 & Point 2 & Point 3     \\
			\hline
			$m_{0}^{U}$          &   2493      & 1523 & 2547      \\
			$M_{1},M_{2},M_{3} $         &   1052,-1255,4497      & 672,-1304,3644.5 & 850.7,373.5,1566.5  \\
			$m_{\tilde{E^c}},m_{\tilde {L}}$      &   150.5,176.1    &  159.9,217 &566.4,698.1  \\
			$m_{H_{u}},m_{H_{d}}$           &    777.1,3817     & 883, 2437 & 221.9,397.5    \\
			$m_{\tilde{Q}},m_{\tilde{U^{c}},m_{\tilde{D^{c}}}}$    & 2324.1,3283.3,3285.6 & 1404.8,1979.4,1984.3 & 2336.6,3255.5,3288.2 \\
			$A_{t}=A_{b},A_{\tau}$            &    -5912,-836.5     & -5990,-1088  & -5542,466.4      \\
			$\tan\beta$                      & 56.8 & 59.4  &54.7  \\
			\hline
			$m_h$            &  125    & 125 & 124         \\
			$m_H$            &  2845    & 2843 & 2366       \\
			$m_{A} $         &  2827     &  2824 & 2350      \\
			$m_{H^{\pm}}$    &  2847    & 2844 & 2368        \\
			\hline
			$\Delta a_{\mu}$
			& \textcolor{red}{18.86$\times 10^{-10}$} & \textcolor{red}{26.98$\times 10^{-10}$} & \textcolor{red}{19.06$\times 10^{-10}$}
			\\			
            \hline			
			$m_{\tilde{\chi}^0_{1,2}}$
			& 450,1160 & 283, 1180 & 303,365\\
			$m_{\tilde{\chi}^0_{3,4}}$
			& 4878,4879 & 4322,4383 & 3546,3546     \\
			$m_{\tilde{\chi}^{\pm}_{1,2}}$
			&1164,4879  & 1185,4383 & 304, 3526 \\
			\hline
			$m_{\tilde{g}}$  & 9001    &  7358 & 3472     \\
			\hline $m_{ \tilde{u}_{L,R}}$
			& 7793,8308  & 6462,6583 &3726,4300      \\
			$m_{\tilde{t}_{1,2}}$
			& 6503,6760  & 5060,5423 & 2573, 3129    \\
			\hline $m_{ \tilde{d}_{L,R}}$
			& 7984,8320 & 6463,6591 &  3727,4411\\
			$m_{\tilde{b}_{1,2}}$
			& 6718,7532 & 5385,5845 & 2635, 3777\\
			\hline
			$m_{\tilde{\nu}_{1}}$
			& 725       & 816 & 385     \\
			$m_{\tilde{\nu}_{3}}$
			& 897       & 918  & 579    \\
			\hline
			$m_{ \tilde{e}_{L,R}}$
			& 749,491   & 842,301 & 334, 1083  \\
			$m_{\tilde{\tau}_{1,2}}$
			& 490,999    & 299,969 & 399,1201  \\
			
			\hline
			$\sigma_{SI}(pb)$
			& 2.01$\times 10^{-14}$ & 2.26$\times 10^{-14}$ & 1.03$\times 10^{-12}$
			\\
			$\Omega_{CDM}h^2$
			& 0.042      & 0.004  & 0.007
    \\
			\hline
			\hline
		\end{tabular}
	}
	\caption{ The sparticle and Higgs masses (in GeV units) for the benchmark points.
		\label{table1}}
\end{table}
\textbf{Collider Searches: Bounds and Prospects.--} The SUSY has been searched inclusively at the LHC. 
The light EW sector of the EWSUSY model that is used to address the $\Delta a_\mu$ will have been excluded by those searches, except that the spectrum is compressed or the mass scale is high. 
All benchmark points are featured by large mass splitting between the left-handed and right-handed sleptons,. 
For the benchmark points 1 and 2, the left-handed sleptons are relatively heavy and the right-handed sleptons have masses close to the LSP (it is opposite for benchmark point 3), such that the direct slepton search at 13 TeV 139 fb$^{-1}$~\cite{Aad:2019vnb,Aad:2019qnd} cannot probe them. 
The most sensitive search for those two points is the trilepton search~\cite{Aaboud:2018jiw} for Wino production $\tilde{\chi}^0_2 \tilde{\chi}^\pm_1 \to (\tilde{\ell} \ell) (\tilde{\ell} \nu) \to (\tilde{\chi}^0_1 \ell \ell) (\tilde{\chi}^0_1 \ell \nu)$. Those two benchmark points just fall beyond the current bound for this channel and may be probed/excluded in the near future, when the same analysis is performed on higher integrated luminosity dataset.  
For benchmark point 3, although the $\tilde{\chi}^0_2$ has relatively large mass spitting with the LSP and decays through $\tilde{\chi}^0_2 \to \tilde{\ell} \ell \to \tilde{\chi}^0_1 \ell \ell$ producing leptons in the final state, it is Bino dominate and its production rate at the LHC is suppressed. To probe the light compressed wino for this benchmark point, the HE-LHC with collision energy 27 TeV is required~\cite{CidVidal:2018eel}.

\textbf{Discussion and Conclusion.--} 
Combining the BNL and Fermi-Lab experimental results for the measurements of
 muon anomalous magnetic moment, we have $4.2 \sigma$ deviation from the SM,
 which strongly suggests the new physics around 1 TeV.
 We analyzed the corresponding five Feynman diagrams
in the SSMs, and showed that the EWSUSY is definitely needed. 
We realized the EWSUSY in the MSSM with the GmSUGRA. 
We found large viable parameter space, which is consistent
with all the current experimental constraints. In particular, the
LSP neutralino can be at least as heavy as 550 GeV.
Most of the viable parameter space can be probed at the future HL-LHC,
while we do need the future HE-LHC to probe some viable parameter space.  
The benchmark points, which have been studied in this work, are based 
on the assumption of R-parity conservation. 
If the R-parity is broken via the operators $U_i^c D_j^c D_k^c $ in the superpotential, 
the corresponding collider bounds may become much weaker and then
the wider classes of benchmark points become viable. 
As a result, testing the EWSUSY which can explain the $\Delta a_\mu$ naturally at the
future HL-LHC and HE-LHC will become a big challenge.

\textbf{Acknowledgments.--} This research was supported by 
the  Projects  11875062  and  11947302  supported  by  
the National  Natural  Science  Foundation  of  China,  
and  by the Key Research Program of Frontier Science, CAS and the Fundamental Research Funds for the Central Universities (TL and JL). WA supported in part by the Grants No. NSFC-11975130, No. NSFC-12035008, by The National Key Research and Development Program of China under Grant No. 2017YFA0402200. The numerical results described in this paper have been obtained via the HPC Cluster of ITP-CAS, Beijing, China.%



\end{document}